\RequirePackage{amsmath}
\documentclass[runningheads]{llncs}

\usepackage{ecir24-sparse-cross-encoder-frame}
\graphicspath{{./figures}}

\begin{document}

\title{Investigating the Effects of\texorpdfstring{\\}{}Sparse Attention on Cross-Encoders}

\author{Ferdinand Schlatt \and Maik Fr{\"o}be \and Matthias Hagen}
\authorrunning{F. Schlatt et al.}
\institute{Friedrich-Schiller-Universit{\"a}t Jena}

\maketitle

\begin{abstract}
  Cross-encoders are effective passage and document re-rankers but less efficient than other neural or classic retrieval models. A few previous studies have applied windowed self-attention to make cross-encoders more efficient. However, these studies did not investigate the potential and limits of different attention patterns or window sizes. We close this gap and systematically analyze how token interactions can be reduced without harming the re-ranking effectiveness. Experimenting with asymmetric attention and different window sizes, we find that the query tokens do not need to attend to the passage or document tokens for effective re-ranking and that very small window sizes suffice. In our experiments, even windows of 4~tokens still yield effectiveness on par with previous cross-encoders while reducing the memory requirements by at least 22\%~/~59\% and being 1\%~/~43\% faster at inference time for passages~/~documents. Our code is publicly available.%
  \footnote{\url{https://github.com/webis-de/ECIR-24}}
\end{abstract}


\section{Introduction}

Pre-trained transformer-based language models~(PLMs) are important components of modern retrieval and re-ranking pipelines as they help to mitigate the vocabulary mismatch problem of lexical systems~\cite{zhang:2022,qu:2021}. Especially cross-encoders are effective~\cite{xiong:2021,macavaney:2019,nogueira:2019,nogueira:2020a} but less efficient than bi-encoders or other classic machine learning-based approaches with respect to inference run time, memory footprint, and energy consumption~\cite{scells:2022}. The run time issue is particularly problematic for practical applications as searchers often expect results after a few hundred milliseconds~\cite{arapakis:2014}. To increase the efficiency but maintain the effectiveness of cross-encoders, previous studies have, for instance, investigated reducing the number of token interactions by applying sparse attention patterns~\cite{sekulic:2020,jiang:2020}.

\begin{figure}[t]
  \includegraphics[width=\textwidth]{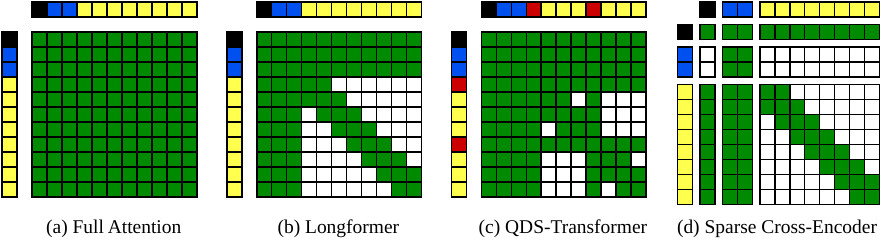}
  \caption{Previous cross-encoder attention patterns~(a, b, and c) and our newly proposed sparse cross-encoder~(d). The marginal boxes denote input tokens (black: [CLS], blue: query, yellow: passage~/~document, red: start of sentence). The inner green boxes indicate token attention. Our new pattern considers the sub-sequences separately (indicated by the added spacing) and is asymmetric.}
  \label{fig-patterns}
  % \vspace{-4ex}
\end{figure}

Sparse attention~PLMs restrict the attention of most tokens to local windows, thereby reducing token interactions and improving efficiency~\cite{tay:2023}. Which tokens have local attention is a task-specific decision. For instance, cross-encoders using the Longformer model~\cite{beltagy:2020} apply normal global attention to query tokens but local attention to document tokens. The underlying idea is that a document token does not require the context of the entire document to determine whether it is relevant to a query---within a document, most token interactions are unnecessary, and a smaller context window suffices
(cf.\ Figure~\ref{fig-patterns}\,(b) for a visualization).

Previously, sparse attention has been applied to cross-encoders to be able to re-rank long documents without cropping or splitting them~\cite{sekulic:2020,jiang:2020}. However, the impact of sparsity on effectiveness has not been investigated in detail. To close this gap, we explore the limits of sparse cross-encoders and try to clarify which token interactions are (un)necessary. As mechanisms to reduce token interactions, we investigate local attention window sizes and disabling attention between sub-sequences (e.g., between the query and the passage or document). Our analyses are based on the following two assumptions.

\Ni~Cross-encoders create contextualized embeddings that encode query and passage or document semantics. We hypothesize that the contextualized embeddings of passage or document tokens do not actually need to encode fine-grained semantics. Rather, an overall ``gist'' in the form of small local context windows is sufficient to estimate relevance well.

\Nii~Cross-encoders allow queries and documents to exchange information via symmetric attention. We hypothesize that full symmetry is unnecessary as we view the query--document relevance relationship as asymmetric: for ranking, it should suffice to determine whether a result is relevant to a query, not vice versa. To further reduce token interactions, we propose a novel configurable asymmetric cross-attention pattern with varying amounts of interaction between~[CLS], query, and passage or document tokens (cf.\ Figure~\ref{fig-patterns}\,(d)).

In experiments on re-ranking tasks from the TREC Deep Learning tracks~\cite{craswell:2019,craswell:2020,craswell:2021,craswell:2022} and the TIREx benchmark~\cite{frobe:2023}, our new model's effectiveness is consistently on par with previous (sparse) cross-encoder models, even with a local window size of only four tokens and asymmetric attention. Further efficiency analyses for sequences with 174~/~4096~tokens show that our sparsification techniques reduce the memory requirements by at least 22\%~/~59\% and yield inference times that are at least 1\%~/~43\% faster.

\section{Related Work}

One strategy for using~PLMs in ranking is to separate the encoding of queries and documents (or passages)~\cite{karpukhin:2020,khattab:2020,santhanam:2022,qu:2021}. Such bi-encoder models are efficient retrievers as the document encodings can be indexed, so only the query needs to be encoded at retrieval time. However, the good efficiency of bi-encoders comes at reduced effectiveness compared to cross-encoders~\cite{hofstatter:2021}, especially in out-of-domain scenarios~\cite{rosa:2022}. Consequently, bi-encoders are often used for first-stage retrieval, while the more effective but less efficient cross-encoders serve as second-stage re-rankers~\cite{xiong:2021,macavaney:2019,nogueira:2019,nogueira:2020a}. We focus on improving the efficiency of cross-encoders while maintaining their effectiveness.

One strategy to make cross-encoders more efficient is to reduce the model size via knowledge distillation~\cite{hofstatter:2020,hofstatter:2021a,lin:2021}. During knowledge distillation, a smaller and more efficient model aims to mimic a larger ``teacher'' model. The distilled models can often match the effectiveness of the teacher model at only a fraction of the computational costs~\cite{wang:2021}, indicating that PLMs can be ``overparameterized'' for re-ranking~\cite{hofstatter:2021}. We follow a similar idea and try to reduce token interactions in cross-encoders using sparse attention techniques to substantially lower computational costs at comparable effectiveness.

Previously, sparse attention~PLMs aimed to increase the processable input length~\cite{tay:2023}. Instead of full attention across the entire input, a sparse PLM restricts attention to neighboring tokens. For instance, the Sparse Transformer~\cite{child:2019} uses a block-sparse kernel, splitting the input into small blocks where tokens only attend to tokens in their block. Additional strided attention allows for attention between blocks. The Longformer~\cite{beltagy:2020}, BigBird~\cite{zaheer:2020}, and ETC~\cite{ainslie:2020} use a different approach. Tokens can attend to a fixed window of neighboring tokens, with additional global attention tokens that can attend to the entire sequence.

For efficient windowed self-attention, the Longformer, BigBird, and ETC use block-sparse matrices. However, block-sparse techniques often make concessions on the flexibility of window sizes or incur additional overhead that does not affect the resulting predictions. We compare several previously proposed windowed self-attention implementations and find them inefficient in terms of time or space compared to the reduction in operations. Therefore, we implement a custom CUDA~kernel and compare it to other implementations (cf.\ Section~\ref{sec-efficiency-analysis}).

Sparse attention~PLMs have also been applied to document re-ranking. For example, the Longformer without global attention was used as a cross-encoder to re-rank long documents~\cite{sekulic:2020}. However, effectiveness was not convincing as later-appearing document tokens were unable to attend to query tokens. The QDS-Transformer~\cite{jiang:2020} fixed this problem by correctly applying global attention to query tokens and achieved better retrieval effectiveness than previous cross-encoder strategies that split documents~\cite{dai:2019a,yan:2019}. While the QDS-Transformer was evaluated with different window sizes, the effectiveness results were inconclusive. A model fine-tuned on a window size of $64$~tokens was tested with smaller (down to $16$~tokens) and larger window sizes (up to $512$)---always yielding worse effectiveness. We hypothesize that models specifically fine-tuned for smaller window sizes will be as effective as models fine-tuned for larger window sizes.

Besides analyzing models fine-tuned for different window sizes, we hypothesize that token interactions between some input sub-sequences are unnecessary. For example, bi-encoder models show that independent contextualization of query and document tokens can be effective~\cite{karpukhin:2020,khattab:2020,santhanam:2022,qu:2021}. However, the symmetric attention mechanisms of previous sparse PLM~architectures do not accommodate asymmetric attention patterns. We develop a new cross-encoder variant that combines windowed self-attention from sparse~PLMs with asymmetric cross-attention. Cross-attention allows a sequence to attend to an arbitrary other sequence and is commonly used in transformer architectures for machine translation~\cite{vaswani:2017}, computer vision~\cite{petit:2021,feng:2022,sui:2022}, and in multi-modal settings~\cite{ilinykh:2022}.
\section{Sparse Asymmetric Attention Using Cross-Encoders}

We propose a novel sparse asymmetric attention pattern for re-ranking documents (and passages) with cross-encoders. Besides combining existing windowed self-attention and cross-attention ideas, our pattern also flexibly allows for asymmetric query--document interactions (e.g., allowing a document to attend to the query but not vice versa). To this end, we partition the input sequence into the [CLS]~token, query tokens, and document tokens, with customizable attention between these groups and local attention windows around document tokens.

Figure~\ref{fig-patterns} depicts our and previous cross-encoder attention patterns. In full attention, each token can attend to every other token. Instead, Longformer-based cross-encoders apply windowed self-attention to document tokens to which the QDS-Transformer adds global attention tokens per sentence. Our pattern is similar to the Longformer but deactivates attention from query tokens to [CLS] and document tokens. But, [CLS] and document tokens still have access to query tokens. Our hypothesis is that cross-encoders do not need symmetric query--document attention for re-ranking as a one-sided relationship suffices.

\subsection{Preliminaries}

A cross-encoder predicts a relevance score for a query--document pair~$(q, d)$ as follows. Given~$q$ and~$d$ as token sequences $q_1 \dots q_m$ and~$d_{1} \dots d_{n}$ and adopting BERT-style encoding~\cite{devlin:2019}, a concatenation of a special classification token~[CLS], then $q$, then a separator token~[SEP], then $d$, and then another [SEP]~token is passed through a transformer encoder. The output of the last transformer layer is an $s \times h$ matrix~$O$, with $s = m+n+3$ being the total sequence length and $h$~being the embedding dimensionality. Each column of~$O$ is an embedding vector of a token from the input sequence. There is one such $O$-matrix per transformer layer, but only the [CLS]~token embedding of the last transformer layer (first column of that layer's $O$-matrix) is used as input to a final linear transformation that then predicts the relevance score of~$d$ for~$q$.

The transformer encoder internally uses a dot-product attention mechanism~\cite{vaswani:2017}. For a single transformer layer, three separate linear transformations map the embedding matrix~$O'$ of the previous layer to three vector-lookup matrices $Q$,~$K$, and~$V$. An $s \times s$ attention probability matrix that contains the probabilities of a token attending to another is obtained by softmaxing the $\sqrt{h}$-normalized product $Q K^T$. The attention probabilities are then used as weights for the vector-lookup matrix~$V$ to obtain the layer's output embedding matrix:
\[
  O \;=\; \Attention(Q, K, V) \;=\; \softmax\left( \dfrac{QK^T}{\sqrt{h}} \right) V\,.
\]

\subsection{Windowed Self-Attention}

Windowed self-attention was proposed for more efficient sparse PLM~architectures~\cite{beltagy:2020, zaheer:2020, ainslie:2020}. The idea is that a token does not attend to the entire input sequence of length~$s$, but only to a local window of $w$~tokens, with $w \ll s$ (e.g., $w = 4$ means that a token attends to $2 \cdot 4 + 1$~tokens: to the 4~tokens before itself, to itself, and to the 4~tokens after itself). For a window size~$w$, windowed self-attention changes the dot-products of the transformer attention mechanism to windowed variants $\boxdot_w$ and $\odot_w$:
\begin{gather*}
  O \;=\; \Attention_w(Q, K, V) \;=\; \softmax\left( \dfrac{Q \boxdot_w K^T}{\sqrt{h}} \right) \odot_w V\,, \text{ with}\\
  \begin{align*}
    Q_{\textcolor{gray}{(s \times h)}} \boxdot_w K^T_{\textcolor{gray}{(h \times s)}} & \;\rightarrow\; A_{\textcolor{gray}{(s \times 2 w + 1)}}, & \text{ where }a_{i,j} & =  \sum_{l=1}^h q_{i,l} \cdot k_{l,i + j - w - 1}\,,     \\
    P_{\textcolor{gray}{(s \times 2w+1)}} \odot_w V_{\textcolor{gray}{(s \times h)}}  & \;\rightarrow\; O_{\textcolor{gray}{(s \times h)}},       & \text{ where }o_{i,l} & = \sum_{j=1}^{2w+1} p_{i,j} \cdot v_{i + j - w - 1,l}\,.
  \end{align*}
\end{gather*}
Thus, $\boxdot_w$ outputs a band matrix subset of the standard matrix--matrix multiplication, stored in a space-efficient form (non-band entries omitted), and $\odot_w$ multiplies a space-efficiently stored band matrix and a standard matrix. To ensure correctness, we zero-pad windows exceeding the sequence bounds: when either $i + j - w \leq 0$ or $i + j - w > s$, we set $k_{l, i + j - w - 1} = v_{i + j - w - 1, l} = 0$.

For a visual impression of windowed attention, consider the lower right document-to-document attention matrix in Figure~\ref{fig-patterns}\,(d). Only the diagonal band is computed: in Figure~\ref{fig-patterns}\,(d) for $w=1$.

Compared to full self-attention, in theory, windowed self-attention reduces the space complexity from ${\cal{O}}(s^2)$ to ${\cal{O}}(s \cdot (2w+1))$ and the (na{\"i}ve) computational complexity from ${\cal{O}}(s^2 \cdot h)$ to ${\cal{O}}(s \cdot (2w+1) \cdot h)$. However, fully achieving these improvements is difficult in practice. Previous windowed self-attention implementations avoided writing hardware-specific kernels and made concessions regarding flexibility, time efficiency, or space efficiency~\cite{beltagy:2020,zaheer:2020}. Therefore, we implement our own CUDA kernel for windowed self-attention; Section~\ref{sec-efficiency-analysis} compares our implementation's efficiency to previous implementations.

\subsection{Cross-Attention}

Cross-attention is a type of attention where a token sequence does not attend to itself, as in self-attention, but to a different sequence. We use cross-attention to configure attention between different token types. Instead of representing a cross-encoder's input as a single sequence, we split it into three disjoint subsequences: the [CLS] token, the query tokens, and the document tokens (Figure~\ref{fig-patterns}\,(d) visually represents this for our proposed pattern by splitting the marginal vectors; the [SEP] tokens are part of ``their'' respective subsequence). Each subsequence can then have its own individual attention function $\Attention(Q, K, V)$.

We split the vector-lookup matrices column-wise into [CLS], query, and document token-specific submatrices $Q_c$,~$Q_q$,~$Q_d$, etc. These matrices are pre-computed and shared between the different attention functions for efficiency. Restricting attention between subsequences then means to call the attention function for a subsequence's $Q$-matrix and the respective $K$- and $V$-matrices of the attended-to subsequences. For example, to let a document attend to itself and the query, the function call is $\Attention(Q_d, [K_q, K_d], [V_q, V_d])$, where $[\cdot, \cdot]$ denotes matrix concatenation by columns (i.e., $[M, M']$ yields a matrix whose ``left'' columns come from~$M$ and the ``right'' columns from~$M'$).

\subsection{Locally Windowed Cross-Attention}

The above-described cross-attention mechanism using concatenation is not directly applicable in our case, as we want to apply windowed self-attention to document tokens and asymmetric attention to query tokens. Instead of concatenating the matrices~$K$~and~$V$, our mechanism uses tuples~$\mathcal{K}$ and~$\mathcal{V}$ of matrices and a tuple~$\mathcal{W}$ of window sizes to assign different attention window sizes~$w \in \mathcal{W}$ to each attended-to subsequence. As a result, we can combine windowed self-attention with asymmetric attention based on token types. Formally, given $j$-tuples~$\mathcal{K}$ and~$\mathcal{V}$ of matrices~$K_i$ and~$V_i$ and a $j$-tuple $\mathcal{W}$ of window sizes~$w_i$, our generalized windowed cross-attention mechanism works as follows:
\begin{gather*}
  \Attention_\mathcal{W}(Q, \mathcal{K}, \mathcal{V}) \;=\; \sum_{i=1}^{j} P_i \odot_{w_i} V_i\,, \text{ where }                                                                       \\
  \left[ P_1, \ldots, P_j \right] = \softmax \left( \dfrac{\left[ A_1, \ldots, A_j \right]}{\sqrt{h}} \right) \;\,\text{ and }\;\, A_i = Q \boxdot_{w_i} K_i\,.
\end{gather*}

Our proposed attention pattern (visualization in Figure~\ref{fig-patterns}\,(d)) is then formally defined as follows. The [CLS]~token has full attention over all subsequences (Equation~1; for notation convenience, we use $w=\infty$ to refer to full self-attention), the query tokens can only attend to query tokens (Equation~2), and the document tokens can attend to all subsequences but use windowed self-attention on their own subsequence (Equation~3):
\begin{align}
  O_c & = \Attention_{(\infty, \infty, \infty)}(Q_c, (K_c, K_q, K_d), (V_c, V_q, V_d)), \\
  O_q & = \Attention_{(\infty)}(Q_q, (K_q), (V_q)),                                     \\
  O_d & = \Attention_{(\infty, \infty, w)}(Q_d, (K_c, K_q, K_d), (V_c, V_q, V_d)).
\end{align}

\subsection{Experimental Setup} \label{sec:experimental-setup}

We fine-tune various models using the Longformer and our proposed attention pattern with window sizes~$w \in \{\infty, 64, 16, 4, 1, 0\}$ ($\infty$: full self-attention). We start from an already fine-tuned and distilled cross-encoder model%
\footnote{https://huggingface.co/cross-encoder/ms-marco-MiniLM-L-6-v2}
which also serves as our baseline~\cite{reimers:2019}. We additionally fine-tune a QDS-Transformer model with its default $w=64$ window for comparison. All models are fine-tuned for 100,000~steps with 1,000~linear warm-up steps and a batch size of~32 (16~document pairs) with margin MSE loss using MS~MARCO-based knowledge distillation triples~\cite{hofstatter:2021}. For documents, we extend the models fine-tuned on passages using positional interpolation~\cite{chen:2023} and further fine-tune them on document pairs from MS~MARCO Document~\cite{nguyen:2016} for 20,000 steps using RankNet loss~\cite{burges:2010}. Negative documents are sampled from the top $200$~documents retrieved by BM25~\cite{robertson:1994}. We use a learning rate of $7 \cdot 10^{-6}$, an AdamW optimizer~\cite{loshchilov:2019}, and a weight decay of $0.01$. We truncate passages and documents to a maximum sequence length of 512~and~4096 tokens, respectively. All models were implemented in PyTorch~\cite{paszke:2019} and PyTorchLightning~\cite{falcon:2023} and fine-tuned on a single NVIDIA A100 40GB GPU.

We evaluate the models on the TREC~2019--2022 Deep Learning (DL) passage and document retrieval tasks~\cite{craswell:2019, craswell:2020, craswell:2021, craswell:2022} and the TIREx benchmark~\cite{frobe:2023}. For each TREC DL task, we re-rank the top 100~passages~/~documents retrieved by BM25 using \texttt{pyserini}~\cite{lin:2021a}. For TIREx, we use the official first-stage retrieval files retrieved by BM25 and ChatNoir~\cite{potthast:2012,bevendorff:2018} and also re-rank the top 100~documents. We measure nDCG@10 and access all corpora and tasks via \texttt{ir\_datasets}~\cite{macavaney:2021}, using the default text field for passages and documents.

To evaluate time and space efficiency, we generate random data with a query length of $10$~tokens and passage~/~document lengths from $54$~to~$4086$~tokens. For the QDS-Transformer, we set global sentence attention at every 30th~token, corresponding to the average sentence length in MS~MARCO documents. We use the largest possible batch size per model, but up to a maximum of 100.
\section{Empirical Evaluation}

We compare our sparse cross-encoder's re-ranking effectiveness and efficiency to full attention and previous sparse cross-encoder implementations. We also examine the impact of different small window sizes and of our attention deactivation pattern---analyses that provide further insights into how cross-encoders work.

\subsection{Effectiveness Results}
\label{sec-effectiveness-analysis}

\paragraph{In-domain Effectiveness} Table~\ref{tbl-reranking-result} reports the nDCG@10 of various models with different attention patterns and window sizes on the TREC Deep Learning passage and document re-ranking tasks. We group Full Attention and Longformer models into the same category because they have the same pattern but different window sizes in our framework. We fine-tune separate models for passage and document re-ranking (cf.~Section~\ref{sec:experimental-setup}) except models with $w=\infty$. Their lack of efficiency prevents training on long sequences, and we only fine-tune them on passages but include their MaxP scores~\cite{dai:2019a} for documents (in gray).

Since we hypothesize that sparse attention will not substantially affect the re-ranking effectiveness, we test for significant equivalence instead of differences. Therefore, we cannot use the typical t-test, but instead use a paired TOST~procedure (two one-sided t-tests~\cite{schuirmann:1987}; $p < 0.003$, multiple test correction~\cite{lauzon:2009}) to determine if the difference between two models is within $\pm 0.02$. We deem $\pm 0.02$ a reasonable threshold for equivalence since it is approximately the difference between the top two models in the different TREC Deep Learning tasks.

\begin{table}[t]
  \caption{Re-ranking effectiveness as nDCG@10 on TREC Deep Learning~\cite{craswell:2019,craswell:2020,craswell:2021,craswell:2022}. The highest score per task is given in bold. Scores obtained using a MaxP strategy are grayed out. \sigtost~denotes significant equivalence within $\pm 0.02$ (paired TOST~\cite{schuirmann:1987}, $p < 0.003$), compared to Full Attention $w=\infty$ for passage tasks and Longformer $w=64$ for document tasks.}
  \centering
  \footnotesize
  \renewcommand{\tabcolsep}{3pt}
  \begin{tabular}{@{}l@{}lccccccccccccc@{}}
    \toprule
    \multicolumn{2}{l}{\bfseries Task}    & \multicolumn{6}{c}{\bfseries Full Att. / Longformer} & \multicolumn{6}{c}{\bfseries Sparse Cross-Encoder} & \bfseries QDS                                                                                                                                                                                                 \\
    \cmidrule(l@{\tabcolsep}r@{\tabcolsep}){3-8} \cmidrule(l@{\tabcolsep}){9-14} \cmidrule(l@{\tabcolsep}r@{\tabcolsep}){15-15}
    \multicolumn{2}{l}{\hspace{1em} $w=$} & $\infty$                                             & 64                                                 & 16               & 4                & 1                & 0                & $\infty$ & 64                     & 16               & 4             & 1                & 0    & 64                               \\
    \midrule
    \raisebox{-9.5ex}[0em][0em]{\rotatebox{90}{Passage}}
    \multirow[t]{5}{*}{}                  & 2019                                                 & .724                                               & .719\kernSigtost & .725\kernSigtost & .719             & .714             & .694     & .722                   & .717             & .724          & \textbf{.728}    & .715 & .696 & .720\kernSigtost          \\
                                          & 2020                                                 & .674                                               & .681\kernSigtost & .680             & \textbf{.684}    & .676             & .632     & .666                   & .672             & .661          & .665             & .649 & .605 & .682                      \\
                                          & 2021                                                 & \textbf{.656}                                      & .653             & .650             & .645             & .629             & .602     & \textbf{.656}          & .650             & .639          & .647             & .625 & .593 & \textbf{.656}\kernSigtost \\
                                          & 2022                                                 & \textbf{.496}                                      & .494\kernSigtost & .487             & .486             & .481             & .441     & .490                   & .492\kernSigtost & .479          & .484             & .471 & .427 & .495\kernSigtost          \\
    \cmidrule{2-15}
                                          & Avg.                                                 & .619                                               & .619\kernSigtost & .616\kernSigtost & .615\kernSigtost & .607             & .572     & .615\kernSigtost       & .615\kernSigtost & .607          & .612\kernSigtost & .596 & .560 & \textbf{.620}\kernSigtost \\
    \midrule
    \raisebox{-10.5ex}[0em][0em]{\rotatebox{90}{Document}}
    \multirow[t]{5}{*}{}                  & 2019                                                 & \textcolor{gray}{.658}                             & .683             & .678             & .667             & .689             & .663     & \textcolor{gray}{.638} & .672             & .685          & .669             & .692 & .646 & \textbf{.697}             \\
                                          & 2020                                                 & \textcolor{gray}{.622}                             & .640             & .639             & \textbf{.661}    & .655             & .644     & \textcolor{gray}{.636} & .638             & .650          & .642             & .657 & .638 & .639                      \\
                                          & 2021                                                 & \textcolor{gray}{.678}                             & .671             & .681             & \textbf{.683}    & \textbf{.683}    & .629     & \textcolor{gray}{.677} & .681             & .681          & .670             & .679 & .644 & .676                      \\
                                          & 2022                                                 & \textcolor{gray}{.424}                             & .425             & .431             & .425             & .409             & .389     & \textcolor{gray}{.421} & \textbf{.446}    & .443          & .417             & .424 & .405 & .428                      \\
    \cmidrule{2-15}
                                          & Avg.                                                 & \textcolor{gray}{.575}                             & .582             & .586\kernSigtost & .587             & .584\kernSigtost & .556     & \textcolor{gray}{.573} & .590             & \textbf{.594} & .577             & .589 & .561 & .587\kernSigtost          \\
    \bottomrule
  \end{tabular}
  \label{tbl-reranking-result}
  % \vspace{-4ex}
\end{table}

We consider two different reference models for the passage and document re-ranking tasks. The Full Attention cross-encoder has complete information access in the passage re-ranking setting and serves as the reference model for the passage tasks. Since the models without windowed attention ($w=\infty$) only process a limited number of tokens in the document re-ranking setting, we use the standard Longformer ($w=64$) as the reference model for document tasks.

We first examine the effectiveness of the QDS-Transformer. In contrast to the original work~\cite{jiang:2020}, it does not improve re-ranking effectiveness despite having more token interactions. The reference models are statistically equivalent to the QDS-Transformer within $\pm 0.02$ for both passage and document re-ranking.

Next, we examine the effect of independent query contextualization on effectiveness. We compare the reference models for the passage and document tasks with our sparse cross-encoder with window size $w=\infty$~and~$w=64$, respectively. This comparison is a type of ablation test, as the two models being compared have identical configurations, except that our sparse cross-encoders independently contextualize the query. Our model is statistically equivalent within $\pm 0.02$ to the reference model on average across all passage tasks. The same does not hold for the document task, but our sparse cross-encoder is slightly more effective, achieving an $0.008$~higher nDCG@10. We conclude that independent query contextualization only marginally affects re-ranking effectiveness.

Finally, we examine the effect of decreasing window size on effectiveness. For the Longformer, the window sizes~$64$,~$16$,~and~$4$ are all significantly equivalent within $\pm 0.02$ on average across passage tasks. Even reducing the window size to just a single token, meaning a passage token can only attend to its immediate left and right neighboring tokens, reduces effectiveness by only $0.012$ but is no longer statistically equivalent. The results are similar for the document tasks, with the window sizes $16$~and~$1$ being statistically equivalent to the reference model. Furthermore, deactivating attention for passage or document tokens to its own subsequence ($w=0$) does not yield statistically equivalent effectiveness but only drops effectiveness by $0.047$~and~$0.026$~nDCG@10, respectively.

The effect of decreasing window sizes is similar for our sparse cross-encoder model. Window sizes $64$~and~$4$ are statistically equivalent for passage tasks. Window sizes $16$~and~$1$ are not statistically equivalent but only drop effectiveness by $0.012$~and~$0.023$ nDCG@10, respectively. On the document tasks, our sparse cross-encoder models with smaller window sizes are never statistically equivalent within $\pm 0.02$ compared to the reference Longformer with window size~$64$. However, window sizes $64$,~$16$,~and~$1$ slightly improve over the reference model, and window size $4$ is slightly less effective.

In summary, both independent query contextualization and windowed self-attention do not substantially affect re-ranking effectiveness, confirming our initial assumptions. That is, symmetric modeling of the query--passage relationship is unnecessary, and very small window sizes suffice to determine a passage's or document's relevance to the query. Interestingly, we find a window size of~$0$ to still feature competitive effectiveness. In this case, a document (or passage) token cannot attend to other tokens from its sub-sequence, making it similar to a lexical or bag-of-words model. We leave a more in-depth investigation into the implications of these results for future work.

\begin{table}[t]
  \caption{Re-ranking effectiveness as nDCG@10 on TIREx~\cite{frobe:2023}. The average document length per corpus and first-stage (FS) effectiveness are listed for context. We report micro-averaged scores across all queries from a corpus and macro-average these in the ``Average'' row. The highest score per corpus is given in bold. Our sparse cross-encoder models use a window size of 4.}
  \centering
  \footnotesize
  \renewcommand{\tabcolsep}{4.8pt}
  \begin{tabular}{@{}l@{\kern-20pt}rcccccccc@{}}
    \toprule
    \bfseries Corpus                                                                & \bfseries Doc. Len. & \bfseries FS  & \multicolumn{3}{c}{\bfseries monoT5} & \multicolumn{2}{c}{\bfseries monoBERT} & \multicolumn{2}{c}{\bfseries Sparse CE \kern-\tabcolsep}                                                                 \\
    \cmidrule(l@{\tabcolsep}r@{\tabcolsep}){4-6} \cmidrule(l@{\tabcolsep}r@{\tabcolsep}){7-8} \cmidrule(l@{\tabcolsep}){9-10}
                                                                                    &                     &               & Base                                 & Large                                  & 3b                                                       & Base          & Large         & 512           & 4096          \\
    \midrule
    Antique~\tiny{\cite{hashemi:2020}}                                              & 49.9                & .510          & .505                                 & .527                                   & .537                                                     & .507          & .484          & \textbf{.540} & .174          \\
    Args.me~\tiny{\cite{bondarenko:2021,bondarenko:2022}}                           & 435.5               & \textbf{.405} & .305                                 & .338                                   & .392                                                     & .314          & .371          & .313          & .180          \\
    CW09~\tiny{\cite{clarke:2009,clarke:2010,clarke:2011,clarke:2012}}              & 1132.6              & .178          & .186                                 & .182                                   & .201                                                     & .192          & .134          & .198          & \textbf{.212} \\
    CW12~\tiny{\cite{thompson:2013,thompson:2014, bondarenko:2021,bondarenko:2022}} & 5641.7              & \textbf{.364} & .260                                 & .266                                   & .279                                                     & .263          & .251          & .312          & .338          \\
    CORD-19~\tiny{\cite{wang:2020}}                                                 & 3647.7              & .586          & .688                                 & .636                                   & .603                                                     & \textbf{.690} & .625          & .673          & .642          \\
    Cranfield~\tiny{\cite{cleverdon:1967,cleverdon:1991}}                           & 234.8               & .008          & .006                                 & .007                                   & .007                                                     & .006          & .006          & \textbf{.009} & .003          \\
    Disks4+5~\tiny{\cite{voorhees:1998,voorhees:1999, voorhees:1996,voorhees:2004}} & 749.3               & .429          & .516                                 & .548                                   & \textbf{.555}                                            & .514          & .494          & .487          & .293          \\
    GOV~\tiny{\cite{craswell:2002,craswell:2003,craswell:2004}}                     & 2700.5              & .266          & .320                                 & .327                                   & \textbf{.351}                                            & .318          & .292          & .316          & .292          \\
    GOV2~\tiny{\cite{clarke:2004,clarke:2005,buettcher:2006}}                       & 2410.3              & .467          & .486                                 & .513                                   & \textbf{.514}                                            & .489          & .474          & .503          & .460          \\
    MED.~\tiny{\cite{hersh:2004,hersh:2005,roberts:2017,roberts:2018}}              & 309.1               & \textbf{.366} & .264                                 & .318                                   & .350                                                     & .267          & .298          & .237          & .180          \\
    NFCorpus~\tiny{\cite{boteva:2016}}                                              & 364.6               & .268          & .295                                 & .296                                   & \textbf{.308}                                            & .295          & .288          & .284          & .151          \\
    Vaswani                                                                         & 51.3                & .447          & .306                                 & .414                                   & .458                                                     & .321          & \textbf{.476} & .436          & .163          \\
    WaPo                                                                            & 713.0               & .364          & .451                                 & \textbf{.492}                          & .476                                                     & .449          & .438          & .434          & .296          \\
    \midrule
    Average                                                                         & --                  & .358          & .353                                 & .374                                   & \textbf{.387}                                            & .356          & .356          & .365          & .260          \\
    \bottomrule
  \end{tabular}
  \label{tbl-tirex-result}
\end{table}

\paragraph{Out-of-domain Effectiveness} Table~\ref{tbl-tirex-result} reports nDCG@10 on all out-of-domain tasks from the TIREx~\cite{frobe:2023} benchmark of our sparse cross-encoder compared to two other cross-encoders of various sizes: monoT5~\cite{nogueira:2020} and monoBERT~\cite{nogueira:2020a}. Our model uses a window size of $4$~tokens, and all models use a maximum sequence length of $512$~token except for our sparse cross-encoder trained on documents, which has access to a maximum of $4096$~tokens. Overall, the out-of-domain re-ranking effectiveness of all cross-encoders is lower than in-domain. Only the monoT5 large and 3b variants and our sparse cross-encoder trained on passages can improve the ranking of the first-stage retrieval on average across all corpora.

Our model trained solely on passages ($512$-token sequence length) features competitive effectiveness despite having substantially fewer parameters. On average over all corpora, it slightly outperforms both the base and large monoBERT variants and the base variant of monoT5. We emphasize that our model only has about $24$~million parameters, making it around four times smaller than the base variant of monoBERT, nine times smaller than the base variant of monoT5, and fourteen times smaller than monoBERT-large. It is slightly less effective than the monoT5 large variant, and the largest model, monoT5 3b, is the most effective.

In contrast, our model trained on documents is substantially less effective in out-of-domain retrieval. Across all corpora, it is $0.105$~nDCG@10 less effective than our model trained on passages. However, it can take advantage of its longer context length on the corpora containing long documents. For example, on the ClueWeb corpora it is the most effective of all cross-encoder models and features competitive effectiveness on CORD-19, GOV, and GOV2.

\subsection{Efficiency Results}
\label{sec-efficiency-analysis}

Finally, we study how the various attention patterns affect efficiency. We first compare our custom windowed matrix multiplication kernel with previous implementations. We then compare the efficiency of our proposed cross-encoder model to the reference cross-encoder, Longformer, and QDS-Transformer.

\begin{figure}[t]
  \begin{subfigure}{0.495\textwidth}
    \centering
    \includegraphics[width=\columnwidth]{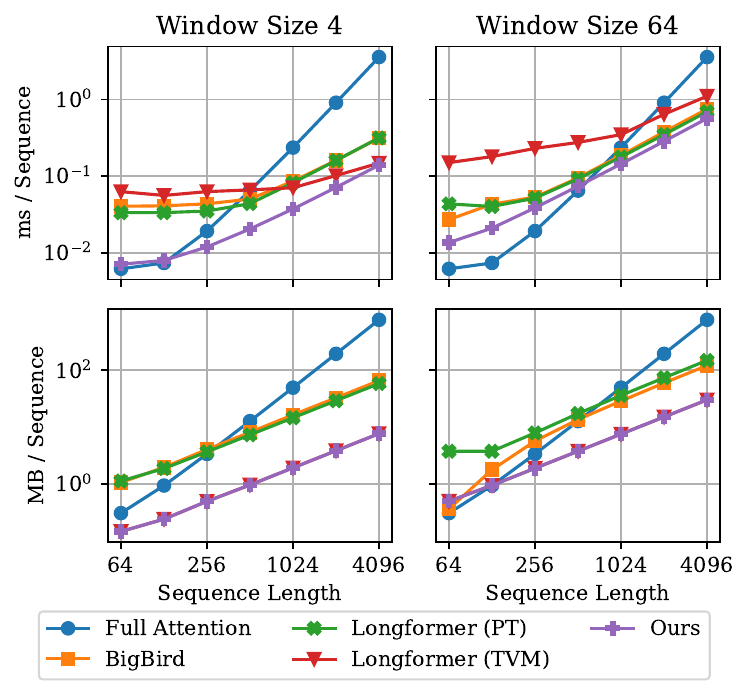}
    \caption{Matrix multiplication kernels.}
    \label{fig-kernel-efficiency}
  \end{subfigure}
  \begin{subfigure}{0.495\textwidth}
    \centering
    \includegraphics[width=\columnwidth]{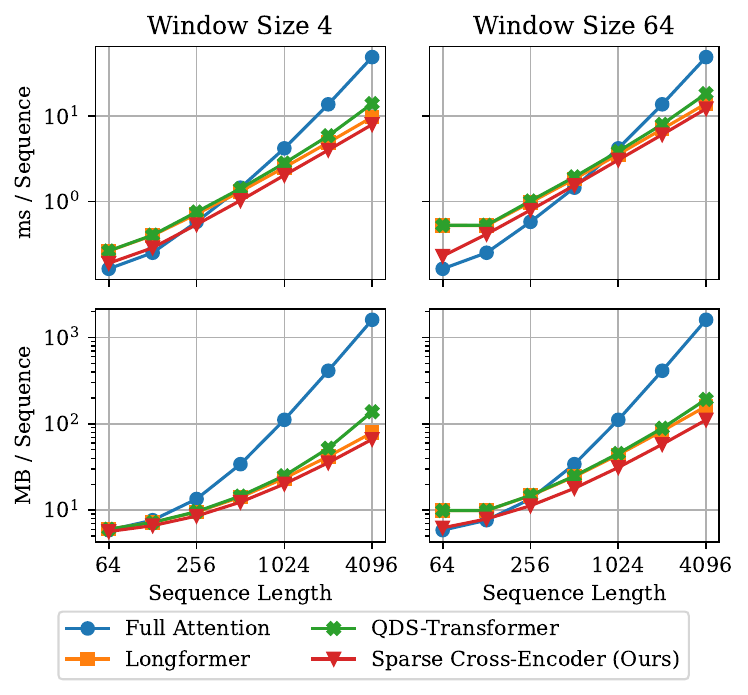}
    \caption{Cross-encoder models.}
    \label{fig-model-efficiency}
  \end{subfigure}
  \caption{Comparison of windowed matrix multiplication kernels~(a) and sparse cross-encoder models~(b) in terms of efficiency. Time~(ms / Document) and space~(GB) efficiency are reported for window sizes $w \in \{4, 64\}$. All plots use a logarithmic scale with base~$2$ on the x-axis and base~$10$ on the y-axis.}
  % \vspace{-4ex}
\end{figure}

\paragraph{Windowed Matrix Multiplication}

Figure~\ref{fig-kernel-efficiency} compares our windowed matrix multiplication kernel with PyTorch's built-in matrix multiplication kernel (Full Attention), Longformer's TVM-based~\cite{chen:2018} implementation, and the two PyTorch-based block-sparse implementations from BigBird and Longformer. All implementations are intended as drop-in replacements for matrix multiplication but use different ways to interface with CUDA and add varying levels of overhead.

We observe similar behavior as \citet{beltagy:2020} for a window size of $64$~tokens. Both block-sparse matrix implementations are time-efficient but sacrifice space efficiency. The opposite is true for the TVM-based kernel. For window size~$4$, the TVM implementation fairs better, achieving similar time efficiency as the PyTorch-based kernels for short sequence lengths and outperforming them at longer sequences. At the same time, it upholds its space efficiency. However, previous kernels are vastly slower compared to full matrix multiplication for shorter sequences. Our custom kernel achieves optimal space efficiency and is faster than all previous windowed matrix multiplication kernels for both window sizes. Compared to full matrix multiplication, our kernel is only slower for the edge case when the window size exceeds the sequence length.

\paragraph{Cross-Encoders Models}

\begin{table}[t]
  \caption{Time and space efficiency of cross-encoder models, including our sparse cross-encoder without our kernel and independent query contextualization. Relative differences to baseline models (underlined) are given in parentheses.}
  \centering
  \footnotesize
  \renewcommand{\tabcolsep}{3pt}
  \begin{tabular}{@{}lccccccc@{}}
    \toprule
    \bfseries Unit       & \bfseries Full Att.       & \bfseries Longf.         & \bfseries QDS.           & \bfseries Sp. CE         & \bfseries Sp. CE        & \bfseries \st{Kernel}   & \bfseries \st{Query}    \\
    \;\,$w=$             & $\infty$                  & 64                       & 64                       & 64                       & 4                       & 4                       & 4                       \\
    \midrule
    \multicolumn{8}{l}{\;\,\textit{Query length 10, Passage length 164}}                                                                                                                                            \\
    \midrule
    \SI{}{\micro\second} & \underline{368}           & 980\,\tiny{$(+166\%)$}   & 995\,\tiny{$(+170\%)$}   & 527\,\tiny{$(+43\%)$}    & 364\,\tiny{$(-1\%)\z$}  & 404\,\tiny{$(+10\%)$}   & 403\,\tiny{$(+10\%)$}   \\
    MB                   & \z\z\underline{9}         & \z15\,\tiny{$(+67\%)\z$} & \z15\,\tiny{$(+67\%)\z$} & \z\z9\,\tiny{$(+0\%)\z$} & \z\z7\,\tiny{$(-22\%)$} & \z\z8\,\tiny{$(-11\%)$} & \z\z7\,\tiny{$(-22\%)$} \\
    \midrule
    \multicolumn{8}{l}{\;\,\textit{Query length 10, Document length 4086}}                                                                                                                                          \\
    \midrule
    ms                   & \z\z49\,\tiny{$(+250\%)$} & \z\underline{14}         & \z18\,\tiny{$(+29\%)$}   & \z12\,\tiny{$(-14\%)$}   & \z8\,\tiny{$(-43\%)$}   & \z9\,\tiny{$(-36\%)$}   & \z8\,\tiny{$(-43\%)$}   \\
    MB                   & 1608\,\tiny{$(+905\%)$}   & \underline{160}          & 192\,\tiny{$(+20\%)$}    & 111\,\tiny{$(-31\%)$}    & 66\,\tiny{$(-59\%)$}    & 84\,\tiny{$(-48\%)$}    & 66\,\tiny{$(-59\%)$}    \\
    \bottomrule
  \end{tabular}
  \label{tbl-efficiency-result}
  % \vspace{-4ex}
\end{table}

Lastly, we compare the efficiency of our sparse cross-encoder model with a standard cross-encoder, the Longformer, and the QDS-Transformer. We use the default implementations from Huggingface~\cite{wolf:2020} but omit BigBird, as it does not support task-specific global attention. Note that the QDS-Transformer is based on the Longformer and uses the same model architecture with a different global attention pattern.

Figure~\ref{fig-model-efficiency} gives a visual overview of the efficiency of the models for windows sizes $4$~and~$64$. Table~\ref{tbl-efficiency-result} reports the time and memory used per sequence for passages of length~$164$ and documents of length~$4086$. These lengths correspond to the average of the longest passage~/~document per top-100 ranking of a TREC Deep Learning query, i.e., the setup simulates re-ranking a batch of $100$~sequences. For ablation analyses, Table~\ref{tbl-efficiency-result} additionally reports the efficiency of our model without our custom kernel and independent query contextualization.

Figure~\ref{fig-model-efficiency} shows our model outperforms the other two sparse cross-encoders for time and space efficiency for both window sizes. The QDS-Transformer is the least efficient due to its additional global sentence attention. The Longformer lies between the QDS-Transformer and our sparse cross-encoder. The efficiency improvements can be attributed to two sources. The first is our improved windowed matrix multiplication kernel, and the second is our cross-attention mechanism. The Listformer uses a similar mechanism but extracts the matrices required for global attention in each transformer layer. Our sparse cross-encoder model splits the sub-sequences once and reuses the extracted matrices for all layers, avoiding repeating the expensive extraction and splitting step.

Table~\ref{tbl-efficiency-result} underlines the efficiency improvements of our model. With a $64$ token window size, our sparse cross-encoder is almost twice as fast and uses $40\%$ less memory than the Longformer on passages. On documents, the difference is less pronounced but still substantial. Our model is $14\%$ faster and uses $31\%$ less memory. However, our sparse cross-encoder achieves the largest efficiency improvements when reducing the window size. Compared to the Longformer with a $64$-token window size, our sparse cross-encoder with a $4$-token window size is $63\%$ faster and uses $53\%$ less memory on passages. On documents, our model is $43\%$ faster and uses $59\%$ less memory. Despite the different window sizes, we deem this a fair comparison because the Longformer was previously not successfully used for re-ranking with smaller window sizes. It acts as the previous sparse cross-encoder efficiency standard.

Ablation tests show that our custom kernel and independent query contextualization both contribute to our model's improved efficiency. Using a Pytorch-based block-sparse windowed matrix multiplication kernel, our model is less time and space-efficient and loses between $9\%$ and $11\%$ percent of its time and space-efficiency improvements. Independent query contextualization only has a marginal effect on space efficiency and a noticeable effect on time efficiency only for passages. The query is generally not long enough compared to passages or documents to substantially impact efficiency in practice.

Comparing our sparse cross-encoder to the standard cross-encoder reveals that there is still room for improvement. Time and space efficiency on documents is orders of magnitude better, and our model uses $22\%$ less memory on passages. But, regarding inference time, our model is on par with the standard cross-encoder for passages. The root cause is that the cross-attention incurs additional overhead. Each sub-sequence uses its own attention function. Multiple smaller attention functions are executed sequentially, while full attention uses a single large attention function for the entire sequence. Recent work on fused-attention kernels~\cite{dao:2023,dao:2022,lefaudeux:2022} has shown that moving the entire attention function to the GPU substantially improves efficiency. At the time of writing, fused-attention kernels do not support asymmetric attention patterns. We leave investigating their applicability to our model to future work.

\section{Conclusion}

We have investigated the impact of sparse attention on the re-ranking effectiveness of cross-encoders by combining windowed self-attention and token-specific cross-attention to analyze \Ni decreasing context sizes for document tokens and \Nii deactivating attention from the query to the [CLS] and document tokens.

In passage and document re-ranking experiments, we find a window size down to four tokens to be as effective as larger window sizes or full attention (significantly equivalent effectiveness within $\pm 0.02$~nDCG@10 compared to previous cross-encoders), and we find that deactivating attention from the query to the [CLS] and document tokens does not impact effectiveness. At the same time, combining the sparsification techniques substantially improves efficiency of passage and document re-ranking. For these efficiency improvements, our custom CUDA kernel and asymmetric cross-attention play substantial roles but the largest gains are achieved using small window sizes.

Sparse attention thus is a viable option for decreasing computational effort without substantially affecting effectiveness. To further increase efficiency, integrating asymmetric cross-attention and windowed self-attention into newly developed fused attention kernels~\cite{dao:2022,dao:2023,lefaudeux:2022} seems to be a promising direction. The flexibility and efficiency of our custom attention pattern also allow for further research into the direction of listwise re-ranking by passing multiple documents to the cross-encoder at once.

\section*{Acknowledgments}
This work has received funding from the European Union's Horizon Europe research and innovation programme under grant agreement No 101070014 (OpenWebSearch.EU, \href{https://doi.org/10.3030/101070014}{https://doi.org/10.3030/101070014}).

\begin{raggedright}
  \small
  \bibliography{ecir24-sparse-cross-encoder-lit}
\end{raggedright}

\end{document}